\documentclass[11pt,a4paper]{article}
\pdfoutput=1
\usepackage{jcappub}

\usepackage[T1]{fontenc}
\usepackage{graphicx}
\usepackage{amsfonts,amsmath,amssymb,tensor}
\usepackage{mathtools}
\usepackage{braket}
\usepackage{bm}
\usepackage{enumerate}

\begin{document}

\title{Enhancement of second-order gravitational waves at Q-ball decay}
\author[a]{Shinta Kasuya,}
\author[b,c]{Masahiro Kawasaki,}
\author[b,c]{and Kai Murai}

\affiliation[a]{Department of Mathematics and Physics, Kanagawa University, Kanagawa 259-1293, Japan}
\affiliation[b]{ICRR, University of Tokyo, Kashiwa, 277-8582, Japan}
\affiliation[c]{Kavli IPMU (WPI), UTIAS, University of Tokyo, Kashiwa, 277-8583, Japan}

\abstract{%
    The recent observation of $^4$He favors a large lepton asymmetry at the big bang nucleosynthesis.
    If Q-balls with a lepton charge decay after the electroweak phase transition, 
    such a large lepton asymmetry can be generated without producing too large baryon asymmetry.
    In this scenario, Q-balls dominate the universe before the decay and induces the sharp transition from the early matter-dominated era to the radiation-dominated era.
    In this transition, the gravitational waves (GWs) are enhanced through a second-order effect of the scalar perturbations.
    We evaluate the density of the produced GWs and show that pulsar timing array observations can probe this scenario depending on the amplitude of the scalar perturbations.
}


\emailAdd{kasuya@kanagawa-u.ac.jp}
\emailAdd{kawasaki@icrr.u-tokyo.ac.jp}
\emailAdd{kmurai@icrr.u-tokyo.ac.jp}

\maketitle

\section{Introduction}
\label{sec: intro}

Recently, the Subaru survey has newly determined the primordial $^4$He abundance $Y_\mathrm{p}$ by observing extremely metal-poor galaxies (EMPGs)~\cite{Matsumoto:2022tlr}.
The reported value is smaller than the previous measurements~\cite{Izotov:2014fga,Aver:2015iza,Hsyu:2020uqb} (see also Ref.~\cite{ParticleDataGroup:2022pth}) and difficult to explain in the framework of the standard big bang nucleosynthesis (BBN).
This anomaly implies the existence of asymmetry in the neutrino sector~\cite{Matsumoto:2022tlr}.
If the degeneracy parameter of electron-neutrinos is as large as $\xi_e \simeq 0.05$, the reported $Y_\mathrm{p}$ can be explained with the predicted deuterium abundance consistent with the observation~\cite{Cooke:2017cwo}.
This value implies the total lepton asymmetry of $\eta_L \simeq 5 \times 10^{-3}$, which is much larger than the observed baryon asymmetry of the universe ($\simeq 10^{-10}$).
Since the sphaleron processes are active in the early universe, the generation of such a large lepton asymmetry is challenging.

One of the mechanisms to generate such a large lepton asymmetry is proposed in Refs.~\cite{Kawasaki:2002hq,Kawasaki:2022hvx}.
In this scenario, the lepton asymmetry is generated through the Affleck-Dine (AD) mechanism~\cite{Affleck:1984fy,Dine:1995kz} and confined in non-topological solitons called Q-balls.
As a result, the produced lepton asymmetry is protected from the sphaleron processes and released at the decay of Q-balls after the electroweak phase transition.

This scenario can also realize the situation that the Q-balls dominate the universe before the decay and then rapidly decrease their energy density through the decay. 
In other words, the Q-balls achieve the early matter-dominated (eMD) era and a sharp transition to the radiation-dominated (RD) era. 
This modifies the evolution of cosmological perturbations such that the scalar perturbations start to oscillate and create a significant amount of gravitational waves (GWs) as second-order effects around the transition~\cite{Inomata:2019zqy,Inomata:2019ivs}.
In this paper, we thus consider the generation of GWs at the Q-ball decay in the scenario of lepton asymmetry generation.
We find that the resultant GWs can be tested in the future PTA observation, the Square Kilometer Array (SKA) depending on the decay temperature of the Q-balls and the amplitude of the primordial curvature perturbations on the corresponding scale.

The generation of GWs at the Q-ball decay was also discussed in Ref.~\cite{White:2021hwi}, which assumes that Q-balls decay instantaneously.
However, since a Q-ball decreases its energy by emitting particles from its surface, Q-ball decay cannot be considered a sudden process.
Consequently, the scalar perturbations are suppressed during the Q-ball decay.
We take this effect into account following the case of primordial black holes studied in Ref.~\cite{Inomata:2020lmk} and evaluate the GW spectrum.

The rest of this paper is organized as follows.
In Sec.~\ref{sec: He and Q-ball}, we shortly review the relation between the recent determination of the $^4$He abundance and the lepton asymmetry, and explain the Q-ball scenario generating the lepton asymmetry.
Production of GWs in the Q-ball scenario is studied in Sec.~\ref{sec: GW}.
We consider the evolution of the scalar perturbations around the eMD era in Sec.~\ref{subsec: scalar} and the generation of the second-order GWs in Sec.~\ref{subsec: 2nd GW}.
In Sec.~\ref{subsec: GW spectrum}, the GW spectrum is shown, and we give prospects for the PTA experiments.
Our results are summarized and discussed in Sec.~\ref{sec: summary and discussion}.

\section{Helium abundance and Q-ball scenario}
\label{sec: He and Q-ball}
Recently, Matsumoto \textsl{et al}.~\cite{Matsumoto:2022tlr} observed 10 EMPGs, in 5 of which $^4$He abundances were reliably derived.
Together with the previously obtained 54 galaxies, the primordial $^4$He abundance was determined as
\begin{align}
    Y_\mathrm{p}
    =
    0.2370^{+0.0034}_{-0.0033}
    \ ,
\end{align}
which is ($\sim 1\sigma$) smaller than the previous measurements~\cite{Izotov:2014fga,Aver:2015iza,Hsyu:2020uqb}.
With the recent measurement of the deuterium abundance~\cite{Cooke:2017cwo}, it is estimated in Ref.~\cite{Matsumoto:2022tlr} that the favored values of the degeneracy parameter of electron neutrinos $\xi_e$ and the effective number of neutrino species $N_\mathrm{eff}$ are
\begin{align}
    \xi_e
    &=
    0.05^{+0.03}_{-0.02}
    \ ,
    \\
    N_\mathrm{eff}
    &=
    3.11^{+0.34}_{-0.31}
    \ ,
\end{align}
respectively.
Since the three flavors of neutrinos have the same amount of asymmetry due to neutrino oscillations, it implies the total lepton asymmetry of
\begin{align}
    \eta_L
    \equiv 
    \frac{n_L}{s}
    \simeq 
    5.3 \times 10^{-3}
    \ ,
\end{align}
where $n_L$ is the total lepton number density and $s$ is the entropy density.
To realize this value of lepton asymmetry, we proposed a scenario utilizing the decay of Q-balls with lepton charge~\cite{Kawasaki:2022hvx}.

\subsection{Q-ball with lepton charge}
\label{subsec: L-ball}

Here, we briefly review the Q-ball scenario.
This scenario is based on the AD mechanism in the minimal supersymmetric standard model (MSSM)~\cite{Affleck:1984fy,Dine:1995kz}.
In the AD mechanism, flat directions with baryon and/or lepton charge start to oscillate after inflation and produce $B-L$ asymmetry.
To discuss the dynamics of the flat directions, we need to specify the potential for the flat directions.
Here, we focus on the gauge-mediated SUSY breaking scenario and consider that a flat direction, called AD field $\phi$, with a lepton charge has a large field value during inflation.\footnote{%
The AD field naturally has a large field value due to the negative Hubble-induced mass term ($\sim c_H H^2 |\phi|^2$ with $c_H$ negative constant) during inflation~\cite{Dine:1995kz}.}
For $|\phi| \gg M_m$, the potential for the AD field is given by
\begin{align}
    V(\phi) 
    &= 
    V_\mathrm{gauge} + V_\mathrm{grav} + V_A
\nonumber\\
    &= 
    M_F^4 \left[\log\left(\frac{|\phi|^2}{M_m^2}\right)\right]^2 
    + m_{3/2}^2 |\phi|^2 \left(1+ K \log\frac{|\phi|^2}{M_*^2}\right)
    + V_A
    \ ,
\end{align}
where $M_m$ is the messenger scale, $M_F$ is the SUSY breaking scale, $m_{3/2}$ is the gravitino mass, $K$ is the coefficient of the one-loop corrections, and $M_*$ is the renormalization scale.
$V_\mathrm{gauge}$ and $V_\mathrm{grav}$ represent the potential lifted by the gauge-mediated and gravity-mediated SUSY breaking effects, respectively.
For $\varphi \equiv |\phi| \gtrsim \varphi_\mathrm{eq}$, $V_\mathrm{grav}$ makes a dominant contribution to $V(\phi)$, while $V_\mathrm{gauge}$ becomes dominant for $\varphi \lesssim \varphi_\mathrm{eq}$.
Here, $\varphi_\mathrm{eq} \simeq \sqrt{2} M_F^2/m_{3/2}$ is the field value where $V_\mathrm{gauge}' \simeq V_\mathrm{grav}'$.
The A-term, $V_A$, breaks the U(1) symmetry and generates the asymmetry.
When the Hubble rate $H$ becomes comparable to the effective mass, the AD field starts to oscillate.
At the same time, the AD field is kicked in the phase direction due to the A-term and, as a result, lepton asymmetry is produced.

In this paper, we assume that the AD field has a field value $\varphi$ much larger than $\varphi_\mathrm{eq}$ during inflation.
Then, the AD field starts to oscillate when $3 H \simeq m_{3/2}$ and the produced lepton asymmetry is given by
\begin{align}
    n_L
    \simeq 
    \epsilon m_{3/2} \varphi_\mathrm{osc}^2
    \ ,
\end{align}
where $\varphi_\mathrm{osc} > \varphi_\mathrm{eq}$ is the value of $\varphi$ when the AD field starts to oscillate, and $\epsilon\, (\leq 1)$ represents the efficiency of the asymmetry generation.
In the following, we assume $\epsilon = 1$ for simplicity.

Next, we review the Q-ball, which can be formed in the AD mechanism.
A Q-ball is a non-topological soliton formed by a complex scalar field, which is stable due to a global U(1) charge~\cite{Coleman:1985ki,Kusenko:1997zq,Kusenko:1997si,Enqvist:1997si,Kasuya:1999wu}.
Once the AD field starts its oscillation, it experiences spatial instabilities and fragments into spherical lumps depending on the shape of the potential.
As a result, almost all the lepton asymmetry carried by the AD field is confined in formed Q-balls.
When $V_\mathrm{grav}$ dominates the potential or equivalently $\varphi \gtrsim \varphi_\mathrm{eq}$, the new-type Q-balls are formed for $K < 0$~\cite{Kasuya:2000sc}.
On the other hand, if $K > 0$, the Q-ball formation is delayed until $\varphi$ becomes around $\varphi_\mathrm{eq}$.
In the following, we focus on the latter type of Q-balls, which is called delayed-type Q-balls~\cite{Kasuya:2001hg}.

The properties of a delayed-type Q-ball are given by~\cite{Hisano:2001dr}
\begin{align}
\begin{aligned}
    Q_\mathrm{init}
    &=
    \beta \left( \frac{ \varphi_\mathrm{eq} }{ M_F } \right)^4
    \ ,
\\
    M_Q
    &=
    \frac{4 \sqrt{2} \pi}{3} \zeta M_F Q^{3/4}
    \ ,
\\
    R_Q 
    &= 
    \frac{1}{\sqrt{2} \zeta} M_F^{-1} Q^{1/4}
    \ ,
\\
    \omega_Q
    &=
    \sqrt{2} \pi \zeta M_F Q^{-1/4}
    \ ,
\end{aligned}
    \label{eq: delayed Q-ball properties}
\end{align}
where $Q_\mathrm{init}$ is the initial value of the Q-ball charge $Q$ when it formed, $M_Q$ is the Q-ball mass, $R_Q$ is the Q-ball radius, and $\omega_Q$ is the frequency of the field oscillation.
$\beta \simeq 6 \times 10^{-4}$~\cite{Kasuya:2001hg} and $\zeta \simeq 2.5$~\cite{Hisano:2001dr} are dimensionless constants.

\subsection{Lepton asymmetry from the decay of Q-balls and their dominance}
\label{subsec: Lepton asymmetry}

Here, we discuss the production of lepton asymmetry from Q-balls.
Q-balls with a lepton charge decay by emitting neutrinos and release their lepton asymmetry.
If the Q-balls decay after the freeze-out of sphaleron processes, the emitted lepton asymmetry is not transformed into baryon asymmetry, and we can thus realize lepton asymmetry much larger than the observed value of the baryon asymmetry of the universe, $\eta_{B,\mathrm{obs}}$.
Although the Q-balls gradually evaporate and emit their lepton asymmetry even before the decay, the abundance of the lepton asymmetry released before the electroweak scale can be smaller than $\eta_{B,\mathrm{obs}}$ for some parameter region.
In this way, the favored value of lepton asymmetry, $\eta_L \simeq 5 \times 10^{-3}$ is obtained without violating cosmological constraints~\cite{Kawasaki:2022hvx}.

If the Q-balls dominate the universe at the decay, the lepton asymmetry is related to the decay temperature as
\begin{equation}
    \eta_L 
    \simeq
    \frac{m_{3/2}\varphi_\mathrm{osc}^2}{4m_{3/2}^2 \varphi_\mathrm{osc}^2/(3 T_\mathrm{dec})}
    =
    \frac{3 T_\mathrm{dec}}{4 m_{3/2}}
    \ ,
    \label{eq: lepton asymmetry for delayed Q-ball}
\end{equation}
which leads to the estimation of the decay temperature as
\begin{align}
    \label{eq:decay_temp_asym}
    T_\mathrm{dec}
    &=
    \frac{4}{3} \eta_L m_{3/2}
\nonumber\\
    &
    \simeq
    3.3~\mathrm{MeV} \,
    \left( \frac{m_{3/2}}{0.5~\mathrm{GeV}} \right)
    \left( \frac{\eta_L}{5 \times 10^{-3}} \right)
    \ .
\end{align}

Let us see whether the Q-balls dominate the energy density of the universe before they decay.
The decay rate is given by
\begin{align}
    \Gamma_Q
    &\equiv 
    - \frac{1}{Q} \frac{\mathrm{d} Q}{\mathrm{d} t}
    \simeq 
    \frac{N_\ell}{Q} \frac{\omega_Q^3}{12\pi^2} 4 \pi R_Q^2
    \simeq
    \frac{\pi^2 N_\ell \zeta}{12 \beta^{5/4}} 
    \frac{m_{3/2}^5}{M_F^4}
    \ ,
\end{align}
where $N_\ell$ is the number of decay channels.
Comparing the decay rate with the Hubble rate, we obtain the decay temperature $T_\mathrm{dec}$ as
\begin{align}
    T_\mathrm{dec} 
    & \simeq 
    \left( \frac{90}{\pi^2 g_*(T_\mathrm{dec})} \right)^{1/4}
    \sqrt{M_\mathrm{Pl}\Gamma_{Q}}
\nonumber\\
    &=
    \left( \frac{90}{\pi^2 g_*(T_\mathrm{dec})} \right)^{1/4}
    \frac{\pi N_\ell^{1/2} \zeta^{1/2}}{2\sqrt{3} \beta^{5/8}} 
    \frac{M_\mathrm{Pl}^{1/2} m_{3/2}^{5/2}}{M_F^2}
\nonumber\\[0.5em]
    &\simeq
    2.69 ~\mathrm{MeV}
\nonumber\\
    & \hspace{14pt} \times 
    \left( \frac{g_*(T_\mathrm{dec})}{10.75} \right)^{-1/4}
    \left( \frac{\beta}{6 \times 10^{-4}} \right)^{-5/8}
    \left( \frac{m_{3/2}}{0.5~\mathrm{GeV}} \right)^{5/2}
    \left( \frac{M_F}{5\times 10^6~\mathrm{GeV}} \right)^{-2}
    \left( \frac{N_\ell}{3} \right)^{1/2}
    \left( \frac{\zeta}{2.5} \right)^{1/2}
    \ ,
    \label{eq:qball_decay_temp}
\end{align}
where $g_*(T_\mathrm{dec})$ is the relativistic degrees of freedom at $T_\mathrm{dec}$.
Since Q-balls behave as non-relativistic matter, they increase their energy fraction and eventually dominate the universe.
The energy density ratio of the Q-balls to radiation at the decay time is given by
\begin{align}
    \left. \frac{\rho_Q}{\rho_\mathrm{r}} \right|_{T_\mathrm{dec}}
    &\simeq
    \frac{ m_{3/2}^2 \varphi_\mathrm{osc}^2 }{ 3 M_\mathrm{Pl}^2 H_\mathrm{osc}^2 }
    \frac{T_\mathrm{R}}{T_\mathrm{dec}}
\nonumber\\
    &\simeq
    9.66 \times 10^6 \,  
    \left( \frac{g_*(T_\mathrm{dec})}{10.75} \right)^{1/4}
    \left( \frac{\beta}{6 \times 10^{-4}} \right)^{5/8}
    \left( \frac{m_{3/2}}{0.5~\mathrm{GeV}} \right)^{-9/2}
    \left( \frac{M_F}{5 \times 10^6~\mathrm{GeV}} \right)^{6}
\nonumber\\
    & \hspace{53pt} \times
    \left( \frac{N_\ell}{3} \right)^{-1/2}
    \left( \frac{\zeta}{2.5} \right)^{-1/2}
    \left( \frac{T_\mathrm{R}}{ 10^5~\mathrm{GeV}} \right)
    \left( \frac{\varphi_\mathrm{osc}}{10^4 \varphi_\mathrm{eq}} \right)^2
    \ ,
    \label{eq: Q-ball energy ratio}
\end{align}
which shows that the Q-balls dominate the universe for the benchmark parameters.
Here, we assumed that the AD field starts to oscillate before reheating, which gives a lower bound on the reheating temperature as
\begin{align}
    T_\mathrm{R} 
    \lesssim 
    T_{\mathrm{R, max}}
    \sim
    \sqrt{m_{3/2} M_\mathrm{Pl}}
    \sim
    10^9~\mathrm{GeV} \, \left( \frac{ m_{3/2} }{0.5~\mathrm{GeV} } \right)^{1/2}
    \ .
    \label{eq: TR upper bound}
\end{align}
Note that the Q-balls should decay before the BBN not to spoil the success of the standard BBN.
Here, we require $T_\mathrm{dec} > 1$~MeV as a constraint.
Therefore, as seen from Eqs.~\eqref{eq:decay_temp_asym} and \eqref{eq:qball_decay_temp}, the Q-ball scenario successfully realizes the universe with a large lepton asymmetry consistent with the recent $^4$He measurement for $T_\mathrm{dec}\sim \mathcal{O}(1)$~MeV and $m_{3/2} \sim \mathcal{O}(0.1)$~GeV.
Moreover, Eq.~\eqref{eq: Q-ball energy ratio} shows that the Q-balls dominate the universe for a long time.

\subsection{Transition from the eMD era to the RD era}
\label{subsec: Q-ball domination}

\begin{figure}[t]
    \centering
    \includegraphics[width=.75\textwidth]{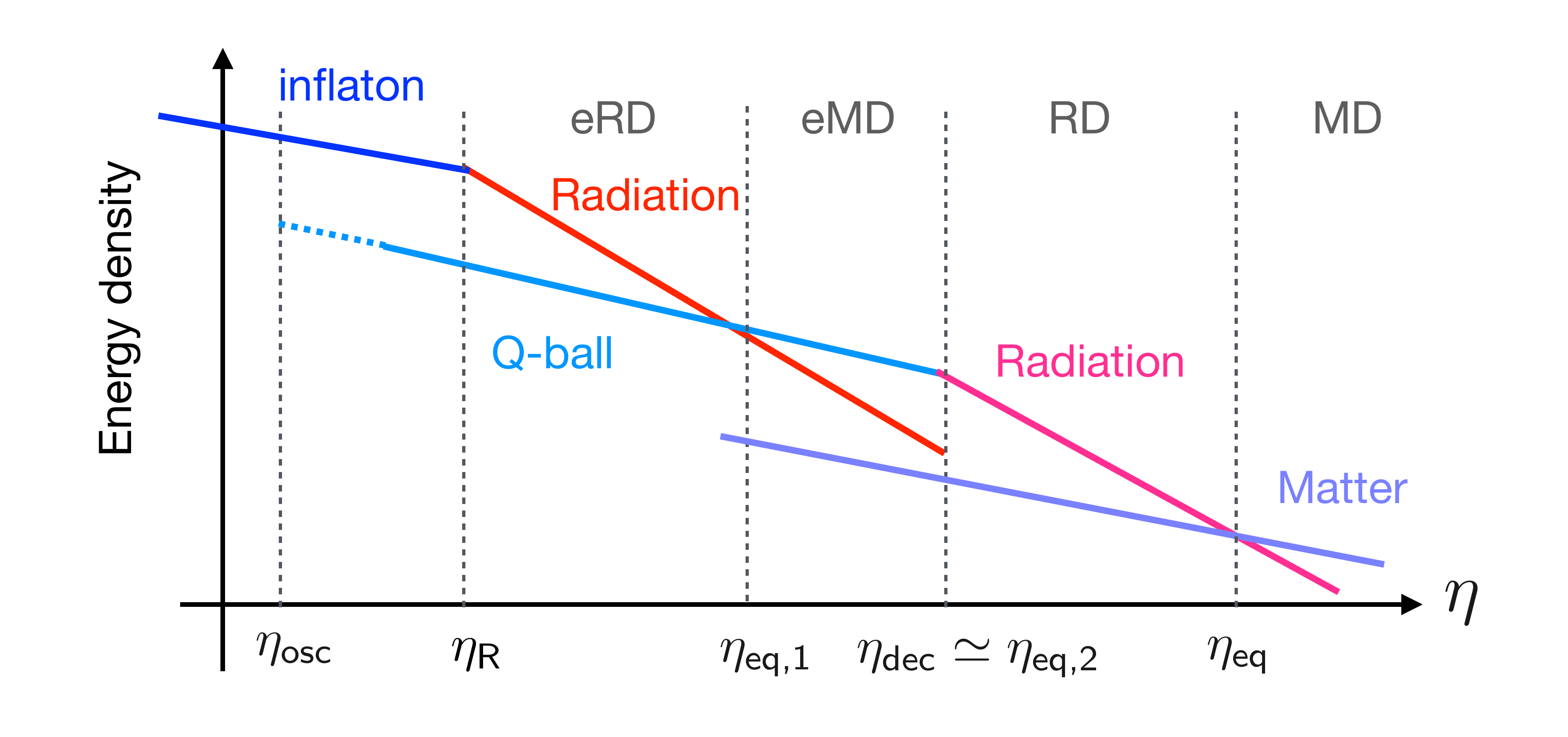}
    \caption{%
        Evolutions of the energy densities of the inflaton, radiation from the inflaton decay, the Q-balls, radiation from the Q-ball decay, and matter.
        $\eta_\mathrm{osc}$ is the conformal time when the AD field starts oscillations.
        $\eta_\mathrm{R}$ represents the reheating epoch.
        $\eta_\mathrm{eq,1}$ and  $\eta_\mathrm{eq,2}$ are the conformal times at matter-radiation equality between radiation and the Q-balls.
        $\eta_\mathrm{eq}$ is the standard matter-radiation equality time.
        }
    \label{fig: Thermal_history}
    \end{figure}

We have seen that the Q-balls can dominate the universe before the decay.
In Fig.~\ref{fig: Thermal_history}, we show the time evolutions of the energy densities of the inflaton, radiation from the inflaton decay, the Q-balls, radiation from the Q-ball decay, and matter ($=$ dark matter $+$ baryons).
Just after the reheating, the universe is dominated by radiation, which we call the early radiation-dominated (eRD) era.
Then, Q-balls increase their energy fraction and have the energy density equal to radiation at $\eta = \eta_\mathrm{eq,1}$, where $\eta$ is the conformal time.
After then, the universe is in the eMD era, which ends at the decay of Q-balls.
We call the equality time around the Q-ball decay $\eta_\mathrm{eq,2}$.
Note that $\eta_\mathrm{eq,2}$ is approximately the same as the time of the Q-ball decay.
The standard RD era follows for $\eta > \eta_\mathrm{eq,2}$.

Thus, the universe experiences the Q-ball-dominated era or eMD era and a transition to the RD era around the Q-ball decay.
Here, we discuss the transition from the eMD era to the RD era, which is important to evaluate the spectrum of the second-order GWs.

\begin{figure}[t]
    \centering
    \includegraphics[width=.6\textwidth]{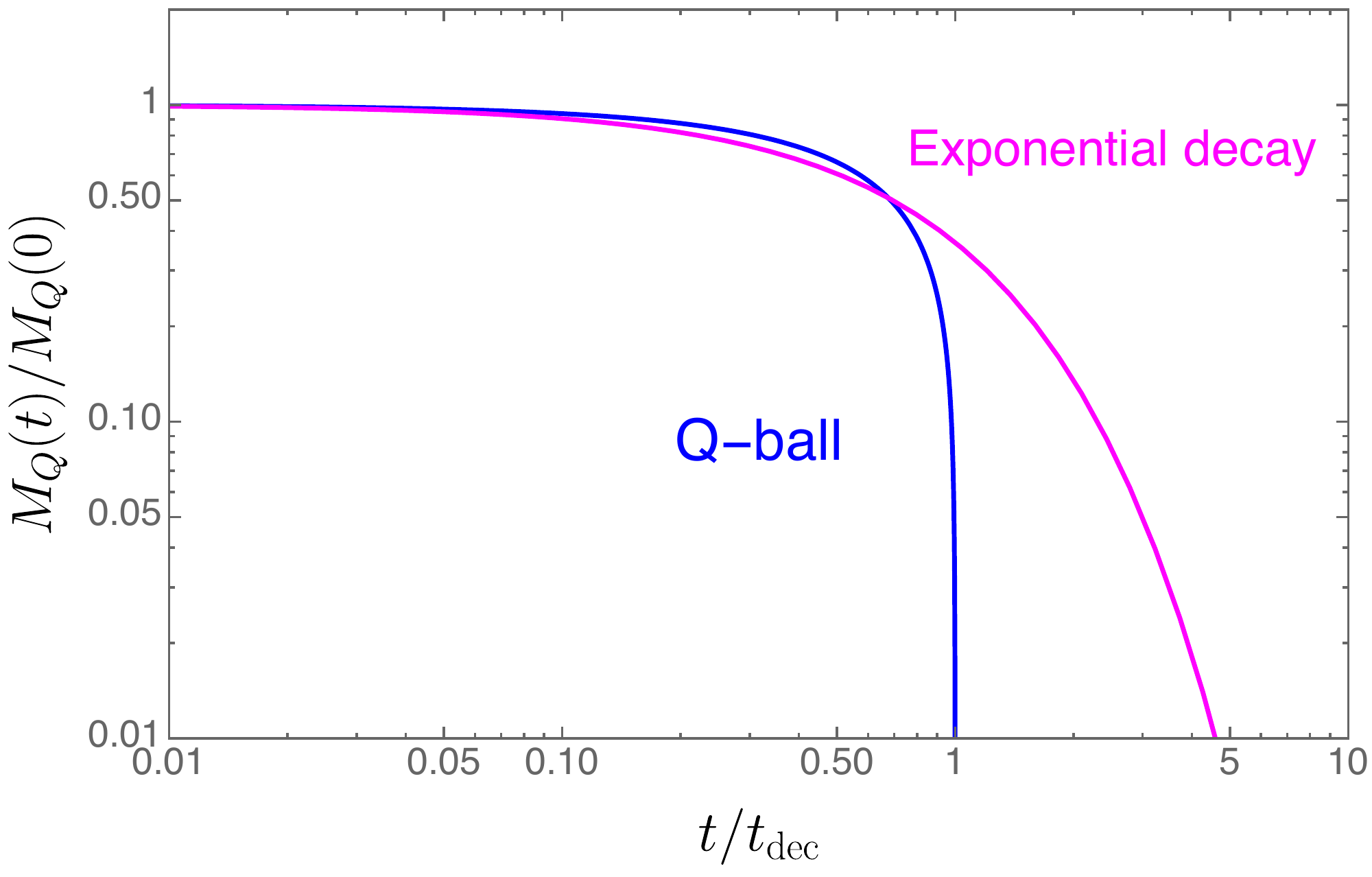}
    \caption{%
        Evolution of a Q-ball mass in the blue line.
        The exponential decay case [$M_Q(t)/M_Q(0) = \exp(-t/t_\mathrm{dec})$] is also shown in the magenta line for comparison.
    }
    \label{fig: Qball_decay}
    \end{figure}

The decay rate of Q-balls can also be written as
\begin{align}
    \Gamma_Q
    = - \frac{1}{Q} \frac{\mathrm{d} Q}{\mathrm{d} t}
    \simeq 
    \frac{N_\ell}{Q} \frac{\omega_Q^3}{12\pi^2} 4 \pi R_Q^2
    =
    \frac{\sqrt{2} \pi^2 \zeta N_\ell M_F}{3} Q^{-5/4}
    \equiv
    \mathcal{A} Q^{-5/4}
    \, .
\end{align}
Thus, the time evolution of the Q-ball charge is given by
\begin{align}
    Q(t) 
    =
    Q_\mathrm{init} 
    \left( 1 - \frac{t}{t_\mathrm{dec}} \right)^{4/5}
    \, ,
\end{align}
where $t$ is the cosmic time satisfying $t = 0$ at the Q-ball formation, and the lifetime $t_\mathrm{dec}$ is defined by
\begin{align}
    t_\mathrm{dec}
    \equiv 
    \frac{4 Q_\mathrm{init}^{5/4}}{5 \mathcal{A}}
    \, .
\end{align}
To discuss the time evolution of the energy density of Q-balls, we consider the time evolution of the Q-ball mass.
Since $M_Q \propto Q^{3/4}$, we obtain
\begin{align}
    M_Q(t) 
    =
    M_{Q} (0)
    \left( 1 - \frac{t}{t_\mathrm{dec}} \right)^{3/5}
    \, .
\end{align}
Thus, the Q-ball decay takes place rapidly compared to the exponential decay.
The time evolution of a Q-ball mass is shown in Fig.~\ref{fig: Qball_decay}, where the evolution is compared with the exponential decay case.

In the eMD and RD eras ($\eta_\mathrm{eq,1} \le \eta < \eta_\mathrm{eq}$), the scale factor and the conformal Hubble rate are given respectively by
\begin{align}
    \frac{a(\eta)}{a(\eta_\mathrm{eq,1})}
    =
    \left\{
        \begin{array}{ll}
            \left( 
                \frac{\eta}{\eta_*}
            \right)^2
            + 
            \frac{2 \eta}{\eta_*}
            \quad&
            (\eta \leq \eta_\mathrm{dec})
        \vspace{3mm} \\ 
            \frac{2 \eta (\eta_\mathrm{dec} + \eta_*) - \eta_\mathrm{dec}^2}
            {\eta_*^2}
            \quad&
            (\eta > \eta_\mathrm{dec})
        \end{array}
    \right.
    \ ,
\end{align}
\begin{align}
    \mathcal{H}(\eta)
    \equiv
    \frac{ 1 }{ a(\eta) }
    \frac{ \mathrm{d} a(\eta) }{ \mathrm{d} \eta }
    =
    \left\{
        \begin{array}{ll}
            \frac{2(\eta + \eta_*)}{\eta^2 + 2 \eta \eta_*}
            \quad&
            (\eta \leq \eta_\mathrm{dec})
        \vspace{3mm} \\ 
            \frac{1}{\eta - \frac{\eta_\mathrm{dec}^2}{2(\eta_\mathrm{dec} +\eta_*)}}
            \quad&
            (\eta > \eta_\mathrm{dec})
        \end{array}
    \right.
    \ ,
\end{align}
where $\eta_* \equiv \eta_\mathrm{eq,1}/(\sqrt{2}-1)$.
Here we assume that the transitions occur suddenly at $\eta_\mathrm{eq,1}$ and $\eta_\mathrm{dec} \simeq \eta_\mathrm{eq,2}$.

Before closing this section, we show the relation between the decay temperature and the corresponding wavelength.
From the entropy conservation, we obtain
\begin{align}
    \frac{a_\mathrm{dec}}{a_\mathrm{eq}}
    =
    \left( 
        \frac{s(T_\mathrm{dec})}{s(T_\mathrm{eq})} 
    \right)^{-1/3}
    =
    \left( 
        \frac{g_{*s} (T_\mathrm{eq})}{g_{*s} (T_\mathrm{dec})}
    \right)^{1/3}
    \frac{T_\mathrm{eq}}{T_\mathrm{dec}}
    \ ,
\end{align}
where $T_\mathrm{eq}$ is the temperature at the most recent matter-radiation equality, and $g_{*s}(T)$ is the relativistic degrees of freedom for entropy density.
On the other hand, the Hubble parameters at the reheating and equality are given by
\begin{align}
    H_\mathrm{dec}^2
    =
    \frac{\pi^2 g_*(T_\mathrm{dec}) T_\mathrm{dec}^4}{90 M_\mathrm{Pl}^2}
    \ ,
    \\
    H_\mathrm{eq}^2
    =
    \frac{2 \pi^2 g_*(T_\mathrm{eq}) T_\mathrm{eq}^4}{90 M_\mathrm{Pl}^2}
    \ .
\end{align}
From these equations, we obtain
\begin{align}
    \frac{a_\mathrm{dec} H_\mathrm{dec}}{a_\mathrm{eq} H_\mathrm{eq}}
    =
    \frac{1}{\sqrt{2}} 
    \left( 
        \frac{g_{*s} (T_\mathrm{eq})}{g_{*s} (T_\mathrm{dec})}
    \right)^{1/3}
    \left( 
        \frac{g_* (T_\mathrm{dec})}{g_* (T_\mathrm{eq})}
    \right)^{1/2}
    \frac{T_\mathrm{dec}}{T_\mathrm{eq}}
    \ .
\end{align}
Thus, the horizon scale at the Q-ball decay is written as
\begin{align}
    k_\mathrm{dec}
    \equiv 
    a_\mathrm{dec} H_\mathrm{dec}
    =
    1.16 \times 10^4 \, \mathrm{Mpc}^{-1}
    \frac{T_\mathrm{dec}}{\mathrm{MeV}}
    \ ,
\end{align}
where we assumed $g_* (T_\mathrm{dec}) = g_{*s} (T_\mathrm{dec}) = 10.75$ and used $g_* (T_\mathrm{eq}) = 2 + 21/4 \times (4/11)^{4/3}$, $g_{*s} (T_\mathrm{eq}) = 43/11$, $a_\mathrm{eq} H_\mathrm{eq} = 0.0103~\mathrm{Mpc}^{-1}$~\cite{Planck:2018vyg}, and $T_\mathrm{eq} = 3400 \times 2.725~\mathrm{K}$. 

\section{Gravitational wave generation in the Q-ball scenario}
\label{sec: GW}

\subsection{Scalar perturbations}
\label{subsec: scalar}

In the conformal Newtonian gauge, the metric perturbations are written as
\begin{align}
    \mathrm{d} s^2
    =
    a^2 \left[ 
        - ( 1 + 2\Phi ) \mathrm{d} \eta^2
        + \left(
            ( 1 - 2\Psi ) \delta_{i j} + \frac{1}{2} h_{i j}
        \right)
        \mathrm{d} x^i \mathrm{d} x^j
    \right]
    \, ,
\end{align}
where $h_{i j}$ is the tensor perturbation satisfying the transverse-traceless condition, $h_i^i = \partial^i h_{i j} = 0$.
In the early universe, we can ignore the anisotropic stress and then obtain $\Phi = \Psi$.
To discuss the generation of the second-order GWs, we focus on the behavior of the transfer function, $\mathcal{T}$, of the gravitational potential $\Phi$ at the transitions around $\eta_\mathrm{eq,1}$ and $\eta_\mathrm{dec}$.

First, we consider the effect of the former transition.
If the perturbations are on superhorizon scales at $\eta_\mathrm{eq,1}$, $\mathcal{T}$ is constant and the amplitude of the perturbations is not suppressed from the primordial one.
On the other hand, if the perturbations are on subhorizon scales at $\eta_\mathrm{eq,1}$, $\mathcal{T}$ during the eMD era is reduced
because the perturbations decay during the eRD era.
The constant value of $\mathcal{T}$ during the eMD era is expressed by a fitting transfer function given by~\cite{Bardeen:1985tr,Inomata:2020lmk}
\begin{align}
    \mathcal{T}_\mathrm{plateau}(x_\mathrm{eq,1})
    \equiv &
    \mathcal{T}(x)|_{\eta_\mathrm{eq,1} \ll \eta \lesssim \eta_\mathrm{dec}}
    \\[0.5em]
    \simeq &
    \frac{\ln[1+0.146 x_\mathrm{eq,1}]}{0.146x_\mathrm{eq,1}}
    \nonumber\\
    & \times 
    \left[ 
        1 + 0.242 x_\mathrm{eq,1}
        + \left( 1.01  x_\mathrm{eq,1}\right)^2
        + \left( 0.341 x_\mathrm{eq,1}\right)^3
        + \left( 0.418 x_\mathrm{eq,1}\right)^4
    \right]^{-1/4}
    \ ,
\end{align}
where $x_\mathrm{eq,1} \equiv k \eta_\mathrm{eq,1}$ and $x \equiv k \eta$.
Note that $\mathcal{T}_\mathrm{plateau}$ is normalized so that $\mathcal{T}_\mathrm{plateau}(x_\mathrm{eq,1} \to 0) \to 1$.

Next, we consider the effect of the latter transition.
During the transition, $\mathcal{T}$ decays following the matter density perturbations at first.
In this phase, $\mathcal{T}$ is approximated as~\cite{Inomata:2019zqy}
\begin{align}
    \frac{\mathcal{T}(t)}{\mathcal{T}_\mathrm{plateau}}
    &\simeq 
    \frac{\rho_\mathrm{m}(t) a^3(t)}{\rho_\mathrm{m}(0) a^3 (0)}
    \simeq
    \frac{M_Q(t)}{M_{Q,\mathrm{init}}}
    \nonumber\\
    &\simeq
    \left( 1 - \frac{t}{t_\mathrm{dec}} \right)^{3/5}
    \ .
\end{align}
This approximation assumes that $\Phi (\propto \mathcal{T})$ is determined only by the matter density fluctuation and requires
\begin{align}
    &3 a^2 |\ddot{\mathcal{T}}|
    \ll
    k^2 \mathcal{T}
    \label{eq: decouple condition}
    \\
    \Leftrightarrow \quad &
    \frac{18}{25 (t_\mathrm{dec} - t)^2 } \ll \frac{k^2}{a^2}
    \ ,
\end{align}
as a necessary condition.
When this condition is violated, $\Phi$ decouples from the matter density perturbations and starts to oscillate.%
\footnote{%
    Even before this condition is violated, $\Phi$ can decouple from the matter density since Eq.~\eqref{eq: decouple condition} is a necessary condition for the approximation. 
    Thus, the violation of this condition gives a lower bound for $S$.
}
The amplitude of $\Phi$ at the onset of oscillations determines the amount of the second-order GWs.
We define the suppression factor of $\mathcal{T}$ at the onset of oscillations by $S$ so that $S(k) = \mathcal{T}_\mathrm{plateau}(x_\mathrm{eq,1})$ in the sudden decay case.
Using the relation $\eta a = 3t$ during the eMD era, we can estimate the decoupling time $\eta_\mathrm{dcpl}$ for a given $k$ as
\begin{align}
    k \eta_\mathrm{dec} - k \eta_\mathrm{dcpl}
    \simeq
    \frac{9\sqrt{2}}{5}
    \ .
\end{align}
As a result, we obtain the lower bound for $S$ as
\begin{align}
    S_\mathrm{low}
    \simeq 
    \left( \frac{9\sqrt{2}}{5 k \eta_\mathrm{dec}}\right)^{3/5}
    \mathcal{T}_\mathrm{plateau}(x_\mathrm{eq,1})
    \ .
    \label{eq: S lower bound}
\end{align}

After the gravitational potential decouples from the matter density fluctuation, the transfer function in the RD era follows the equation of 
\begin{align}
    \mathcal{T}'' + 4 \mathcal{H} \mathcal{T}' + \frac{k^2}{3} \mathcal{T}
    =
    0
    \ ,
\end{align}
where the primes denote the derivatives with respect to $\eta$.
We solve this equation with approximated initial conditions of $\mathcal{T}(x_\mathrm{dec} \equiv k\eta_\mathrm{dec}) = S_\mathrm{low}(k)$ and $\mathcal{T}'(x_\mathrm{dec}) = 0$.
Then, we obtain $\mathcal{T}$ after $\eta_\mathrm{dcpl} \simeq \eta_\mathrm{dec}$ as
\begin{align}
    \mathcal{T}(x)
    =
    S_\mathrm{low}(k) (
        A \mathcal{J}(x) + B \mathcal{Y}(x)
    )
    \ ,
\end{align}
with
\begin{align}
\begin{gathered}
    \mathcal{J}(x)
    =
    \frac{3\sqrt{3} j_1(\frac{x - x_\mathrm{dec}/2}{\sqrt{3}})}{x - x_\mathrm{dec}/2}
    \ , \quad 
    \mathcal{Y}(x)
    =
    \frac{3\sqrt{3} y_1(\frac{x - x_\mathrm{dec}/2}{\sqrt{3}})}{x - x_\mathrm{dec}/2}
    \ ,
    \\[0.5em]
    A =
    \frac{1}{\mathcal{J}(x_\mathrm{dec}) - \frac{\mathcal{Y}(x_\mathrm{dec})}{\mathcal{Y}'(x_\mathrm{dec})} \mathcal{J}'(x_\mathrm{dec})}
    \ , \quad 
    B =
    - \frac{\mathcal{J}'(x_\mathrm{dec})}{\mathcal{Y}'(x_\mathrm{dec})} A
    \ .
\end{gathered}
\end{align}
Here, $j_1$ and $y_1$ are the first and second spherical Bessel functions.

\subsection{Second-order gravitational waves}
\label{subsec: 2nd GW}

Here, we consider the generation of the second-order GWs around the Q-ball decay.
The GW energy density parameter per $\ln k$ is given by
\begin{align}
    \Omega_\mathrm{GW} (\eta, k)
    &\equiv
    \frac{ \rho_\mathrm{GW}(\eta, k) }{ \rho_\mathrm{tot}(\eta) }
    \nonumber\\
    &=
    \frac{1}{24} \left( \frac{ k }{ a(\eta) H(\eta) } \right)^2
    \overline{ \mathcal{P}_h (\eta, k) }
    \ ,
\end{align}
where $\overline{ \mathcal{P}_h (\eta, k) }$ is the time average of the GW power spectrum.
The power spectrum of the second-order GWs is given by~\cite{Kohri:2018awv} (see also Refs.~\cite{Ananda:2006af,Baumann:2007zm,Saito:2008jc,Saito:2009jt,Bugaev:2009zh,Inomata:2016rbd})
\begin{align}
    \overline{ \mathcal{P}_h (\eta, k) }
    =
    4 \int_0^\infty \mathrm{d}v \int_{|1-v|}^{1+v} \mathrm{d}u \,
    \left[
        \frac{ 4v^2 - (1 + v^2 - u^2)^2 }{ 4 v u } 
    \right]^2
    \overline{ I^2(u, v, k, \eta, \eta_\mathrm{dec}) }
    \mathcal{P}_\zeta(u k) \mathcal{P}_\zeta(v k)
    \ ,
\end{align}
where $\mathcal{P}_\zeta$ is the power spectrum of the primordial curvature perturbations.
Note that, during and after the eMD era, $\Phi$ is related to the primordial curvature perturbation $\zeta$ as 
\begin{align}
    \Phi(x)
    =
    \frac{3}{5} \mathcal{T}(x) \zeta
    \ .
\end{align}
$I(u, v, k, \eta, \eta_\mathrm{dec})$ capsulizes the evolution of the scalar perturbations, the GW production, and the GW propagation, and can be expressed as
\begin{align}
    I(u, v, k, \eta, \eta_\mathrm{dec})
    =
    \int_0^x \mathrm{d} \tilde{x} \,
    \frac{ a(\tilde{\eta})}{ a(\eta) } k G_k(\eta, \tilde{\eta})
    f(u, v, \tilde{x}, x_\mathrm{dec})
    \ ,
\end{align}
where $\tilde{x} = k\tilde{\eta}$.
Here, $G_k(\eta, \tilde{\eta})$ is the Green function representing the propagation of GWs induced at $\tilde{\eta}$ and satisfies
\begin{align}
    G_k''(\eta, \tilde{\eta})
    + \left( k^2 - \frac{a''(\eta)}{a(\eta)} \right) G_k(\eta, \tilde{\eta})
    =
    \delta(\eta - \tilde{\eta})
    \ ,
\end{align}
where the primes denote the derivatives with respect to $\eta$, not $\tilde{\eta}$.
$f(u, v, \tilde{x}, x_\mathrm{dec})$ corresponds to the second-order source of the GWs including the transfer of the scalar perturbations and is expressed as
\begin{align}
    & f(u, v, \tilde{x}, x_\mathrm{dec})
    \nonumber\\
    & \equiv
    \frac{
        3
        \left[
            2 ( 5 + 3w(\tilde{\eta}) ) \mathcal{T}(u \tilde{x}) \mathcal{T}(v \tilde{x}) 
            + 4 \mathcal{H}^{-1} 
            ( \mathcal{T}'(u \tilde{x}) \mathcal{T}(v \tilde{x}) 
            + \mathcal{T}(u \tilde{x}) \mathcal{T}'(v \tilde{x}) )
            + 4 \mathcal{H}^{-2} \mathcal{T}'(u \tilde{x}) \mathcal{T}'(v \tilde{x}) 
        \right]
    }
    {25 ( 1 + w(\tilde{\eta}) )}
    \ ,
\end{align}
where $w(\tilde{\eta}) \equiv p/\rho$ is the equation-of-state parameter at $\tilde{\eta}$, and $\mathcal{T}(x)$ is an abbreviated notation of $\mathcal{T}(x, x_\mathrm{eq,1}, x_\mathrm{dec})$.
Note that the primes denote derivatives with respect to the conformal time as mentioned above, and thus $\mathcal{T}'(u \tilde{x}) = \partial \mathcal{T}(u \tilde{x})/\partial \tilde{\eta} = u k \partial \mathcal{T}(u \tilde{x})/\partial (u \tilde{x})$.

Next, we estimate the GW amplitude.
We assume that the primordial power spectrum of the curvature perturbations is given by the scale-invariant spectrum with a cut-off of the following form as
\begin{align}
    \mathcal{P}_\zeta (k)
    =
    C^2 A_\mathrm{s} \Theta(k_\mathrm{NL} - k)
    \ .
\end{align}
Here, $k_\mathrm{NL}$ is the cut-off scale where the matter perturbations $\delta_\mathrm{m}$ becomes unity at the Q-ball decay. 
$A_\mathrm{s} = 2.1 \times 10^{-9}$ is the amplitude on the CMB pivot scale $k_* = 0.05~\mathrm{Mpc}^{-1}$~\cite{Planck:2018vyg}, while a constant $C$ represents the difference from $A_\mathrm{s}$, so that the amplitude is equal to $A_\mathrm{s}$ for $C=1$.
Since we are interested in scales much smaller than the CMB scale, $\mathcal{P}_\zeta$ does not necessarily have the amplitude the same as that on the CMB scale, $C$ could be much larger than unity on smaller scales.

The non-linear scale $k_\mathrm{NL}$ is determined as follows.
During the eMD era, the matter perturbation is related to $\mathcal{T}$ through Poisson's equation on subhorizon scales as
\begin{align}
    \frac{9}{10} k^2 \mathcal{T}_\mathrm{plateau} (x_\mathrm{eq,1}) \phi_k
    \simeq 
    \frac{3}{2} \mathcal{H}^2 \delta_\mathrm{m}
    \ ,
    \label{eq: non-linear condition}
\end{align}
where $\phi_k$ is the constant amplitude of the gravitational potential during the eRD era and related to the curvature perturbation as $|\phi_k| \simeq 2|\zeta|/3 \simeq 2 \mathcal{P}_\zeta^{1/2}/3$.
Note that the factor $9/10$ expresses the evolution of $\Phi$ during the transition from the eRD era to the eMD era.
Solving this equation with the condition $\delta_\mathrm{m} = 1$, we obtain $k_\mathrm{NL} \eta_\mathrm{dec} \to x_{\mathrm{dec, NL}}^{(\mathrm{lim})} \simeq 467$ for $C = 1$ in the limit of $\mathcal{T}_\mathrm{plateau}(x_\mathrm{eq,1}) \to 1$.
We can approximate $k_\mathrm{NL} \eta_\mathrm{dec} \simeq x_{\mathrm{dec, NL}}^{(\mathrm{lim})}$ if $\eta_\mathrm{dec}/\eta_\mathrm{eq,1}$ is much larger than $x_{\mathrm{dec, NL}}^{(\mathrm{lim})}$.
This is because all the relevant scales are superhorizon at $\eta_\mathrm{eq,1}$ if the duration of the eMD era is sufficiently long.
For larger $C$, the nonlinear scale $k_\mathrm{NL}$ or $x_\mathrm{dec,NL}$ becomes smaller due to larger $|\phi_k|$.

To evaluate the power spectrum of the generated GWs, we consider the conformal time $\eta_c$ after the induced GWs become constant and before the late-time equality.
Due to the resonance between the oscillations of $\mathcal{T}$'s in $f$ and the Green function $G_k$, $\overline{\mathcal{P}_h}$ has a steep peak with a cut-off around $k_\mathrm{NL}$.
The peak of the GW spectrum at $\eta_c$ is approximated by~\cite{Inomata:2019ivs,Inomata:2020lmk}
\begin{align}
    \Omega_{\mathrm{GW,RD}}^{(\mathrm{res})} (\eta_c, k)
    \simeq &
    2.9 \times 10^{-7} Y 
    F(n_\mathrm{s,eff})
    C^4 A_s^2 S_\mathrm{low}^4(k) (k \eta_\mathrm{dec})^7
    \nonumber \\
    &\times \frac{s_0(k)}{8}
    \left( 
        15 - 10 s_0^2(k) + 3 s_0^4(k)
    \right)
    \ ,
    \label{eq: 2nd GW formula}
\end{align}
where $Y \simeq 2.3$ is the numerical factor and $F(n_\mathrm{s,eff})$ is defined as
\begin{align}
    F(n_\mathrm{s,eff})
    &\equiv
    \int_{-1}^1 \mathrm{d}s \, (1-s^2)^2
    \left( \frac{3-s^2}{4} \right)^{n_\mathrm{s,eff}-1}
    \nonumber \\
    &=
    \frac{ 3^{n_\mathrm{s,eff}-1} 4^{2-n_\mathrm{s,eff}} }
    { n_\mathrm{s,eff}(3+2n_\mathrm{s,eff}) }
    \left[
        \left(\frac{2}{3}\right)^{n_\mathrm{s,eff}} (n_\mathrm{s,eff}-3)
        +(n_\mathrm{s,eff}^2 - n_\mathrm{s,eff} + 3)
        _2F_1\left(
            \frac{1}{2}, -n_\mathrm{s,eff}; \frac{3}{2};\frac{1}{3}
        \right)
    \right]
    \ ,
\end{align}
where $_2F_1$ is hypergeometric function.
Here, $n_\mathrm{s,eff}$ represents the tilt of the scalar perturbations including the suppression effects due to the transitions, $S_\mathrm{low}(k)$.
In our setup, since the duration of the eMD era is sufficiently long as can be seen from Eq.~\eqref{eq: Q-ball energy ratio}, we can consider $x_\mathrm{eq,1} \ll 1$ and $\mathcal{T}_\mathrm{plateau}(x_\mathrm{eq,1}) \simeq 1$ for all $k$'s of interest.
Thus, we take $n_\mathrm{s,eff} = 1-6/5$ considering Eq.~\eqref{eq: S lower bound}.
For this value of $n_\mathrm{eff}$, we obtain $F(n_\mathrm{s,eff} = 1-6/5) \simeq 1.61$.%
\footnote{%
    If the duration of the eMD era is short and $x_\mathrm{eq,1} \gg 1$ for $k$'s of interest, $\mathcal{T}_\mathrm{plateau}$ is proportional to $k^{-2}$ and thus we should use $n_\mathrm{s,eff} = 1 - 6/5 - 4$ and $F(n_\mathrm{s,eff} = 1-6/5-4) \simeq 6.58$ instead.
}
The function of $s_0$ in the second line of Eq.~\eqref{eq: 2nd GW formula} reflects the cut-off of $\mathcal{P}_\zeta$ at $k_\mathrm{NL}$, and $s_0$ is defined by
\begin{align}
    s_0(k)
    =
    \left\{
        \begin{array}{ll}
            1
            \quad&
            (k \leq \frac{2}{1 + \sqrt{3}} k_\mathrm{NL})
        \vspace{3mm} \\ 
            2 \frac{k_\mathrm{NL}}{k} - \sqrt{3}
            \quad&
            (\frac{2}{1 + \sqrt{3}} k_\mathrm{NL} \leq k \leq \frac{2}{\sqrt{3}} k_\mathrm{NL})
        \vspace{3mm} \\ 
            0
            \quad&
            (k \geq \frac{2}{\sqrt{3}} k_\mathrm{NL})
        \end{array}
    \right.
    \ .
\end{align}

The energy density parameter at the present time is given by~\cite{Ando:2018qdb}
\begin{align}
    \Omega_\mathrm{GW}(\eta_0, k) h^2
    =
    0.83 \left( \frac{g_{*, c}}{10.75} \right)^{-1/3}
    \Omega_{\mathrm{r}, 0} h^2 
    \Omega_{\mathrm{GW,RD}}^{(\mathrm{res})} (\eta_c, k)
    \ ,
\end{align}
where $g_{*, c}$ is the effective degrees of freedom at $\eta_c$, and $\Omega_{\mathrm{r},0}$ is the current density parameter of radiation.
In the following, we use $g_{*,c} = 10.75$ and $\Omega_{\mathrm{r},0} h^2 = 4.2 \times 10^{-5}$.

\subsection{Gravitational wave spectrum}
\label{subsec: GW spectrum}

\begin{figure}[t]
    \centering
    \includegraphics[width=.8\textwidth]{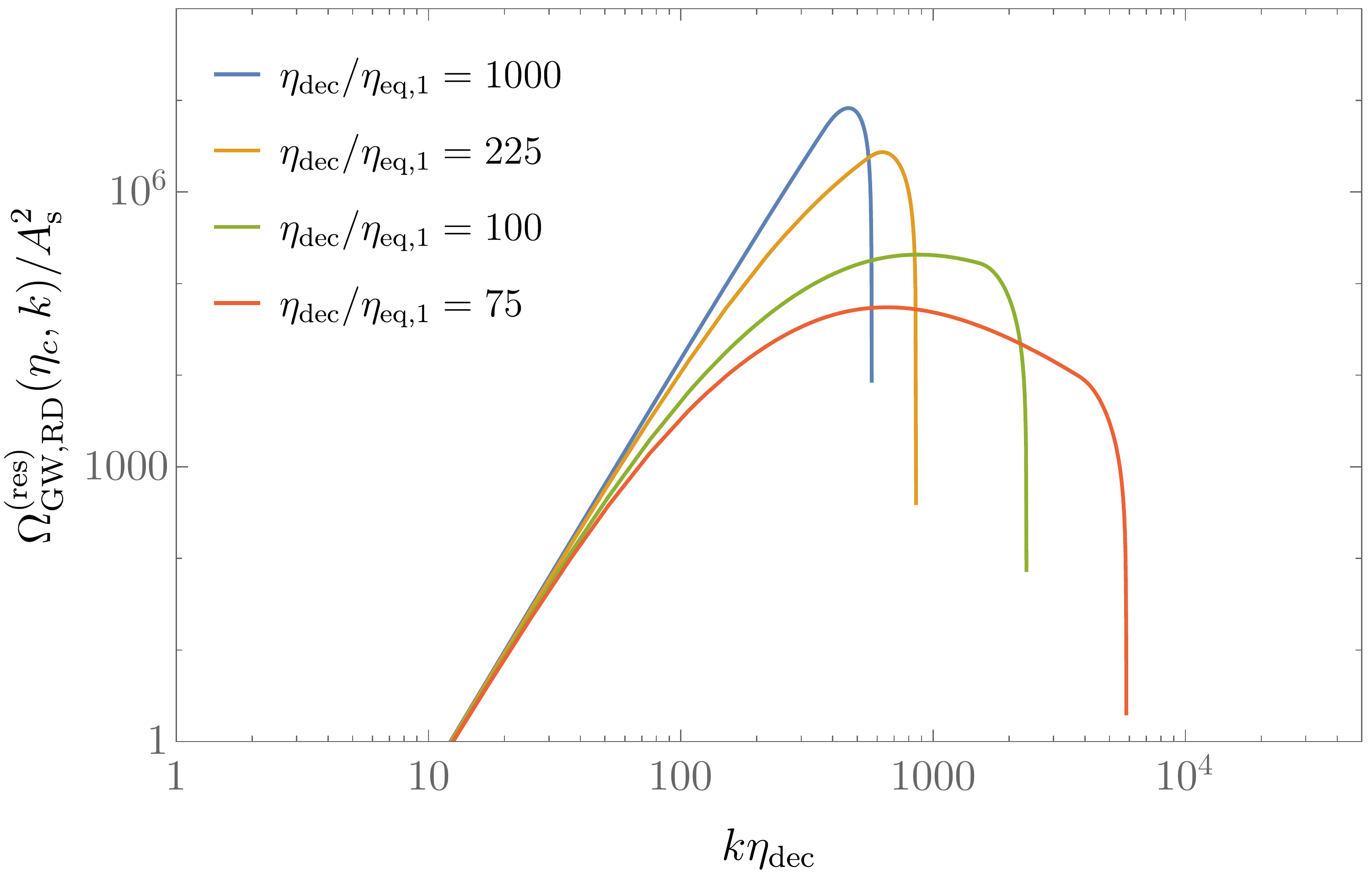}
    \caption{%
        Shape of the GW spectrum for different values of $\eta_\mathrm{dec}/\eta_\mathrm{eq,1}$.
        Here, we fix $C = 1$.
    }
    \label{fig: GW shape}
    \end{figure}

Now, we evaluate and discuss the power spectrum of the GWs generated by the second-order effects around the Q-ball decay.
First, we show the shape of the GW abundance for $C = 1$ and different values of $\eta_\mathrm{dec}/\eta_\mathrm{eq,1}$ in Fig.~\ref{fig: GW shape}.
The cut-off scale depends on $\eta_\mathrm{dec}/\eta_\mathrm{eq,1}$ through the determination of $k_\mathrm{NL}$.
For smaller $\eta_\mathrm{dec}/\eta_\mathrm{eq,1}$, $\mathcal{T}_\mathrm{plateau}$ is suppressed and thus the cut-off scale becomes larger while the GW amplitude becomes smaller.
The value of $C$ also affects the cutoff scale through Eq.~\eqref{eq: non-linear condition}.
Although the choice of $n_\mathrm{s,eff} = 1 - 6/5$ is invalid for smaller $\eta_\mathrm{dec} / \eta_\mathrm{eq, 1}$, we fix $n_\mathrm{s,eff} = 1 - 6/5$ in Fig.~\ref{fig: GW shape}, since $F(n_\mathrm{s,eff})$ does not change the shape of the GW spectrum but affects the overall factor.

\begin{figure}[t]
    \centering
    \includegraphics[width=.8\textwidth]{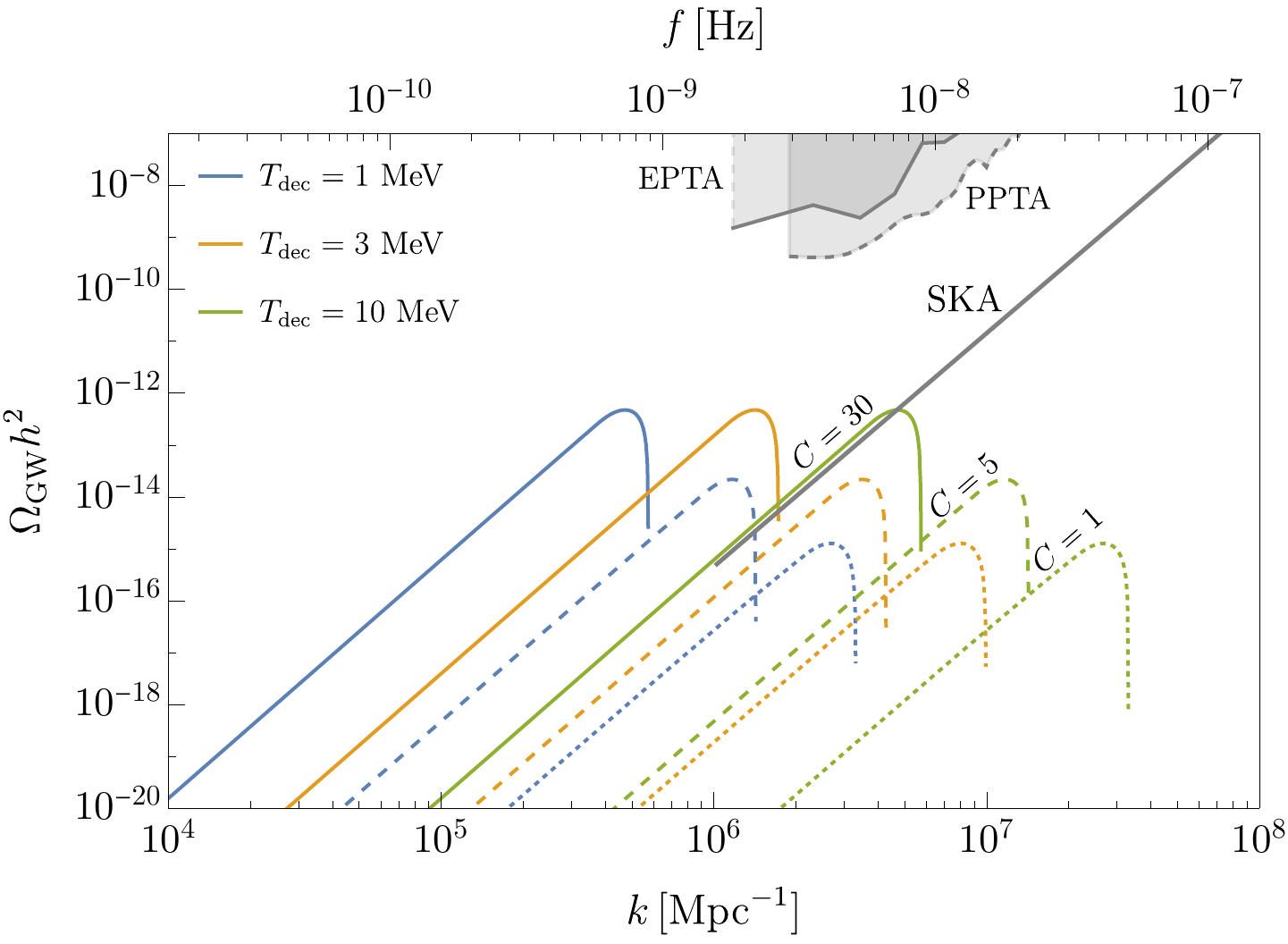}
    \caption{%
        Current density parameter of the GWs for $T_\mathrm{dec} = 1, 3, 10$~MeV from left to right and $C = 1, 5, 30$ from bottom to top.
        We fix $\eta_\mathrm{dec}/\eta_\mathrm{eq,1} = 1000$.
        The gray-shaded regions are constrained by EPTA~\cite{Lentati:2015qwp} and PPTA~\cite{Shannon:2015ect}.
        The gray line shows the future sensitivity of SKA~\cite{Janssen:2014dka}, which is evaluated following Ref.~\cite{Inomata:2018epa}.
        }
    \label{fig: GW vs PTA}
\end{figure}

Next, we show the current density parameter of the GWs for $T_\mathrm{dec} = 1, 3, 10$~MeV and $C = 1, 5, 30$ in Fig.~\ref{fig: GW vs PTA}.
We fix $\eta_\mathrm{dec}/\eta_\mathrm{eq,1} = 1000$ since we are interested in the case where the eMD era is sufficiently long as implied in Eq.~\eqref{eq: Q-ball energy ratio}.
The gray-shaded regions are the constraints by EPTA~\cite{Lentati:2015qwp} and PPTA~\cite{Shannon:2015ect}.
The gray line shows the future sensitivity of SKA~\cite{Janssen:2014dka}, which is evaluated following Ref.~\cite{Inomata:2018epa}.
As shown in Fig.~\ref{fig: GW shape}, the amplitude of the GWs is determined only by $k \eta_\mathrm{dec}$ for fixed values of $\eta_\mathrm{dec}/\eta_\mathrm{eq,1}$ and $C$.
Thus, the difference in the decay temperature does not affect the amplitude but the peak scale, which can be seen in Fig.~\ref{fig: GW vs PTA}.
On the other hand, the difference in $C$ affects both the amplitude and the peak scale.
While the GW spectrum for fixed $k$ scales as $C^4$ through Eq.~\eqref{eq: 2nd GW formula}, the position of the peak depends on $C$ through the determination of $x_\mathrm{dec, NL}$.
For $T_\mathrm{dec} \lesssim 10$~MeV, the GW can be detected in SKA depending on the value of $C$.


\section{Summary and Discussion}
\label{sec: summary and discussion}

In this paper, we have considered a scenario explaining the newly determined value of $Y_\mathrm{p}$ through the Q-balls with a lepton charge.
In this scenario, the Q-balls dominate the universe and then decay emitting lepton asymmetry.
Due to the transition from the eMD era to the RD era at the Q-ball decay, the scalar perturbations can produce a large amount of GWs as the second-order effect.
The power spectrum of the produced GWs depends on various quantities related to the evolution of the background cosmology and the scalar perturbations.
Among them, the duration of the eMD era and the time dependence of the decay process are largely fixed in the scenario of interest.
Thus, we evaluated the resonant production of GWs by taking the amplitude of the primordial curvature perturbations and the Q-ball decay temperature as free parameters.
Consequently, we found that the resultant GW spectrum can have a peak around $k = \mathcal{O}(10^6)$~Mpc$^{-1}$ with an amplitude of $\Omega_\mathrm{GW} h^2 \gtrsim \mathcal{O}(10^{-13})$, which is detectable by SKA.

Although we have focused on the resonant contribution to the GWs, there are non-resonant contributions.
Here, we comment that the non-resonant contribution to the GWs does not affect our discussion on the detectability in SKA.
This is because the non-resonant contribution is dominant for smaller $k$ with the abundance smaller than the peak of the resonant contribution by $\lesssim \mathcal{O}(10^{-3})$, as shown in Refs.~\cite{Inomata:2019ivs,Inomata:2020lmk}.

This mechanism of the GW production at Q-ball decay was also discussed in Ref.~\cite{White:2021hwi}.
They assume that the Q-balls instantaneously decay and do not take into account the suppression due to the gradual transition from the eMD era to the RD era.
Since they consider a type of Q-balls that decays slower than the delayed-type Q-balls, the GW spectrum that they obtained will be suppressed more severely than in our case.
A similar discussion can also be applied to the gravity-mediated type of Q-balls.

\begin{acknowledgments}
We would like to thank Keisuke Inomata for useful discussion on their work.
This work was supported by JSPS KAKENHI Grant Nos. 20H05851(M.K.), 21K03567(M.K.), JP20J20248 (K.M.) and World Premier International Research Center Initiative (WPI Initiative), MEXT, Japan (M.K., K.M.).
K.M. was supported by the Program of Excellence in Photon Science.
\end{acknowledgments}

\small
\bibliographystyle{JHEP}
\bibliography{Ref}

\end{document}